%% file: lni-author-template.tex
\documentclass[english]{lni}

\usepackage{tcolorbox}
\usepackage{fancyvrb}
\usepackage{xcolor}
\usepackage{subcaption}

\begin{document}

\title[Meta-Solving DSL]{A Domain-Specific Language for Formulating Hybrid Quantum-Classical Meta-Solver Strategies}

 \author[1]{Nick Poser}{nick.poser@student.kit.edu}{0009-0001-1915-4419}
 \author[1]{Domenik Eichhorn}{domenik.eichhorn@kit.edu}{0000-0001-9428-024X}
 \author[1]{Ina Schaefer}{ina.schaefer@kit.edu}{0000-0002-7153-761X}
 \affil[1]{Karlsruhe Institute of Technology, Germany}
\maketitle

\begin{abstract}
A key challenge when designing hybrid quantum–classical workflows is the identification of quantum candidates, that is, determining for which specific problems quantum advantages could potentially be achieved.
When choosing between several candidates, it is crucial to consider the characteristics specific to the problem, as these can fundamentally determine how successful quantum or classical approaches will be.
This implies that specialized expertise is needed to use hybrid quantum-classical workflows successfully.
To address this challenge, we propose a domain-specific language (DSL) to express best-practices in solution strategies using a universal representation that is easy to use and share.
This DSL provides a flexible approach to design hybrid quantum-classical workflows and to automate decisions for a wide range of problems, supporting decisions down to problem-specific details while remaining technically independent.
Furthermore, we propose a framework that is built around our DSL that enables the execution of defined workflows using the ProvideQ toolbox as an orchestration layer.
All contributions from this publication are open source.
\end{abstract}

\input{sections/01_introduction}
\input{sections/02_background_relatedwork}
\input{sections/03_concept}
\input{sections/04_implementation}
\input{sections/05_case_study}
\input{sections/06_conclusion}

\printbibliography

\end{document}

%% file: sections/01_introduction.tex
\section{Introduction}

Although current quantum computing is limited by noisy intermediate-scale quantum (NISQ) devices, its theoretical ability to achieve up to superpolynomial speedups on well-structured problems~\cite{aaronson2022much} already calls for preparing its use in high-performance applications~\cite{moller2017impact}.
Realizing this potential is widely expected to require effective integration with established classical computing approaches~\cite{humble2021quantum}.
Hybrid quantum-classical workflows have been proposed to model these new complex systems~\cite{cranganore_paving_2024}.

An essential consideration in the design of hybrid quantum-classical workflows is the identification of quantum candidates, i.e., tasks that could potentially gain an advantage from quantum techniques compared to classical ones.
This step is difficult because the decision between quantum and classical methods must account for the particular characteristics of the problem, its intended application, and the capabilities of both quantum and classical hardware.
As a result, the selection process is highly context-dependent and typically requires expert knowledge~\cite{dalzell2023quantum}.
Although approaches for quantum candidate identification exist, the current knowledge about best-practices is still very limited, which makes the implementation of efficient hybrid quantum-classical workflows a challenging task~\cite{desdentado2025quantum}.

In this paper, we address this challenge by introducing a Domain-Specific Language (DSL) extension for the ProvideQ toolbox~\cite{eichhorn_provideq_2025}.
Our goal is to offer a new, easy-to-read and easy-to-share format for specifying hybrid quantum-classical workflows.
This extension supports the specification of hybrid quantum-classical workflows through an interface that aims to be more maintainable and scalable than graph-based representations, while remaining more accessible and comprehensible than low-level APIs.
Within our DSL, we explicitly focus on identifying quantum candidates using Meta-Solver Strategies~\cite{eichhorn_hybrid_2024}, which break down problem-solving processes into several subroutines and then exploit problem-specific features to assess, based on our expert knowledge, whether quantum computing can be beneficially integrated into the problem-solving pipeline.

In addition to specifying solution strategies, we developed a complete execution environment around our DSL framework that simplifies running hybrid quantum-classical workflows.
Here, the ProvideQ toolbox~\cite{eichhorn_provideq_2025} supplies problem definitions and an orchestration layer to execute solution strategies via interchangeable backends, enabling us to flexibly respond to ongoing progress in quantum algorithms and hardware.
Contributions from this paper are available open-source on GitHub: \href{https://github.com/ProvideQ/meta-solver-strategy-lang}{https://github.com/ProvideQ/meta-solver-strategy-lang}.

%% file: sections/02_background_relatedwork.tex
\section{Related Work and Background}

This section provides background on hybrid quantum–classical workflows and on quantum software frameworks that can be used to implement them. Alongside this, we outline the current state of the art in both areas and position our work in relation to existing approaches.

\subsection{Hybrid Quantum-Classical Workflows}

While quantum computers are expected to provide up to superpolynomial speedups, these performance increases are usually only possible in very problem-specific scenarios with well-structured problems~\cite{aaronson_how_2022}.
Hybrid quantum-classical workflows try to utilize these potential quantum speedups in combination with existing classical computing architectures to leverage the best performance of every domain~\cite{phillipson2023classification}.
An implementation of a hybrid quantum-classical workflow can go from low-level hybrid algorithms such as QAOA~\cite{farhi_quantum_2014} or VQE~\cite{kandala2017hardware} that combine quantum circuits with classical optimizers, to complex high-performance computing applications that aim to orchestrate large computing workflows between different kinds of quantum and classical backends~\cite{chen2024multi, pehlivanoglu2026qurator, chundury2025scaling}. 
Applications are often seen, but not restricted to combinatorial optimization problems~\cite{ellinas2024hybrid, smierzchalski2024hybrid, lopez2026hybrid} or material simulation~\cite{yao2021gutzwiller, ye2024hybrid}.

To reduce the complexity and create standards to implement such hybrid quantum-classical workflows, several works proposed frameworks to model them. 
Weder et al. introduced QuantME~\cite{weder2020integrating}, which is a modeling extension of the graphical Business Process Modeling Language (BPML), adding the explicit expression of quantum computing tasks into workflows;
Sivarajah et al. propose a tool called Tierkreis~\cite{sivarajah2022tierkreis}, a tool enables the programming of hybrid quantum-classical algorithms through dataflow graphs;
Cranganore et al.~\cite{cranganore_paving_2024} describe how existing Workflow Management Systems (WMS) can be extended with quantum subroutines with a strong focus on scientific applications;
De Maio et al. introduced the Rigoletto language~\cite{de_maio_rigoletto_2024}, which can define hybrid quantum-classical workflows via a JSON format;
and Raubenolt and Blankenberg introduced the GalaxyQ platform~\cite{raubenolt2025galaxyq}, which provides tooling for a graph-based implementation of hybrid-quantum classical workflows with a strong focus on reproducibility.

With the exception of Rigoletto~\cite{de_maio_rigoletto_2024}, all previously discussed works employ a graphical, graph-oriented paradigm to represent workflows. A comparable visual strategy is also adopted in our own work on Hybrid Meta-Solving~\cite{eichhorn_hybrid_2024}, a divide-and-conquer approach for modeling hybrid quantum-classical workflows that focuses on solver configuration.  
In contrast to these earlier efforts, we are the first to pursue a workflow specification and execution mechanism that is based on a Domain-Specific Language. Our objective is to leverage the benefits of DSLs in combination with our Hybrid Meta-Solver approach to improve scalability and maintainability~\cite{klint2010impact}. 
Furthermore, our DSL-centered design enables us to incorporate conditions for quantum candidate selection directly into the workflow by encoding decisions based on problem-specific characteristics.

\subsection{Quantum Software Frameworks for Workflow Implementation}

Quantum software tools that are suitable to implement hybrid quantum-classical workflows include low-level libraries such as Qiskit~\cite{Qiskit}, Qrisp~\cite{seidel_qrisp_2022}, or Pennylane~\cite{bergholm2022pennylane}. These Python-based toolkits provide fundamental interfaces to a wide range of quantum algorithms, allowing users to augment their classical workflows with quantum components.
High-level frameworks that add further abstraction layers and are thus more accessible include Qoro Quantum's Divi~\cite{qoro_divi}, which automatically handles optimizing, parallelizing, and distributing jobs across different types of classical and quantum hardware; 
the QuAST Decision Tree~\cite{poggel2024creating}, which automates the solution of optimization problems using NISQ algorithms and provides a recommended workflow for how to apply these algorithms; 
Qiskit’s functions tool~\cite{qiskit_functions}, which streamlines algorithm exploration and application development by abstracting parts of the quantum software development workflow; 
and the Kipu Quantum Hub~\cite{kipu_quantum_hub}, which exposes quantum solvers via a JSON API and enables their use through a BPMN-based workflow engine.

Another high-level framework is the ProvideQ toolbox~\cite{eichhorn_provideq_2025}. In contrast to the previously discussed tools, it is explicitly designed around the Meta-Solving paradigm and already supports configuring solvers for various optimization problems, as well as running them on multiple quantum simulators and on real Kipu Quantum backends.  
With the work presented in this paper, we augment the ProvideQ toolbox with a DSL, which will simplify the deployment and execution of complex Meta-Solver strategies.

%% file: sections/03_concept.tex

\section{Language Design}

The language we propose is a Domain-Specific Language (DSL)~\cite{voelter2013dsl} designed to express Meta-Solver Strategies.
A Meta-Solver Strategy is a hybrid quantum-classical workflow that decomposes a problem into multiple sub-problems.
Problems can be solved using different approaches, and the decision for each approach is based on problem characteristics.
Our DSL is designed around familiar programming concepts, targeting users with an academic or computer science background.

\subsection{Why create a DSL?}
DSLs trade general-purpose expressiveness for domain focus.
Researchers like Kosar et al. \cite{kosar2012program} have shown that developers understand DSL programs better and faster than those in general-purpose languages.
A key advantage of DSLs is that they hide implementation complexity, allowing us to focus on the domain itself and describe problems and solutions without needing to understand the underlying low-level details.
This abstraction supports technical independence, making DSLs more robust to changes in underlying technology.
Technical challenges like these can be handled by an underlying framework, while the language itself remains stable and unaffected by changes.
Additionally, even the best language-specific API that abstracts away most technical details still binds users to a specific programming language and technology stack.
A DSL, on the other hand, can be designed to be a transferable format that can be used to persist and exchange information.
Any framework that can interpret the DSL can execute it, independently of the surrounding technology stack.
Having one technology-independent artifact also simplifies comparison and benchmarking.
For our topic to express hybrid quantum-classical workflows, technical independence is particularly important because the field of quantum computing is rapidly evolving, resulting in technical challenges such as unstable APIs, changing libraries, and shifting standards.
The only dependency that remains is on the external solvers and their specific parameters.
However, this is unavoidable for any language that needs to express solver calls.
Additionally, a transferable format helps to create a language that can be universally used and understood by the community and that can be executed by different frameworks and technology stacks.

\subsection{Language Concepts}

To express Meta-Solver Strategies, the DSL needs to model \textit{problem types} with their concrete \textit{problem instances}.
We represent these using a type/instance model similar to typed object-oriented programming languages, keeping the syntax familiar and intuitive.
We define problem types and problem instances as follows:

\begin{tcolorbox}[fonttitle=\bfseries, title=Problem Type, size=small]
A problem type defines a certain problem which can have concrete instances that can be solved.
An example for a problem type is the Traveling Salesman Problem (TSP), which defines the problem of finding the shortest possible route that visits a set of cities and returns to the origin city.
\end{tcolorbox}

\begin{tcolorbox}[fonttitle=\bfseries, title=Problem Instance, size=small]
A problem instance is a concrete instance of a problem type, which is defined in a standardized format.
Problem instances have properties, such as the number of variables or the number of constraints.
\end{tcolorbox}

As we want to solve problems using Meta-Solver Strategies, we also need to express the intent of solving a problem instance in the language using the keyword \texttt{solve}.
Consider this running example, which we will extend to a full Meta-Solver Strategy in this section:

\definecolor{darkgreen}{RGB}{0, 128, 0}
\definecolor{darkblue}{RGB}{0, 16, 128}
\definecolor{darkred}{RGB}{163, 21, 21}
\begin{Verbatim}[commandchars=\\\{\}]
\textcolor{violet}{solve} \textcolor{teal}{VRP} \textcolor{darkblue}{vrp}:
  \textcolor{darkgreen}{// Now solve something}
\end{Verbatim}

This sketch also shows key syntax choices: colons start new blocks, which are indented to improve readability.
In the example, uppercase \texttt{VRP} denotes the problem type, while lowercase \texttt{vrp} denotes a concrete instance.

Now that we're able to express that we want to solve a certain problem instance, we also need to express how we want to solve it.
A \textit{solver} is called on a problem instance and may require additional parameters (e.g., iteration limits, time limits, or API credentials) to create a \textit{problem solution}.
Parameters in the DSL are passed as key-value pairs to improve readability and robustness against API changes.
Solvers and the resulting problem solution are defined as follows:

\begin{tcolorbox}[fonttitle=\bfseries, title=Problem Solution, size=small]
A problem solution is a answer to a problem instance, defined in a certain format.
The information in a solution can be used to solve other problem instances, like the solution of a TSP clustering problem can be used to solve multiple smaller TSP problems.
\end{tcolorbox}

\begin{tcolorbox}[fonttitle=\bfseries, title=Solver, size=small]
A solver is a service that takes a problem instance in a certain format as input and produces an output.
It produces a solution in a certain format or can transform a problem into another problem or a set of problems.
Solvers can be classical or quantum, and they can differ in their properties, such as their runtime or the quality of their solutions.
\end{tcolorbox}

Since solver calls take certain inputs to produce a solution, they naturally map to method calls in object-oriented programming languages, where the problem instance is the object, and the solver is the method. This results in a syntax similar to a method call:

\begin{Verbatim}[commandchars=\\\{\}]
\textcolor{violet}{solve} \textcolor{teal}{QUBO} \textcolor{darkblue}{qubo}:
  \textcolor{darkblue}{qubo}.\textcolor{brown}{QrispQuboSolver}(
    \textcolor{darkred}{"Max Number of Variables" = "5"})
\end{Verbatim}

So far, the language can express only trivial \textit{Meta-Solver Strategies}, because each strategy only consists of a single solver call.
Since \textit{decisions} are essential for expressing Meta-Solver Strategies, the language needs to be able to express them as well.
These concepts are defined as follows:

\begin{tcolorbox}[fonttitle=\bfseries, title=Decision, size=small]
A decision is a choice that needs to be made in the process of solving a problem instance.
The choice is made between different solvers that all take the same problem instance as input and produce a solution as output.
The best solver for the current context is chosen based on properties of a problem instance and knowledge about the solver's properties.
A decision is made based on a logical formula featuring operations such as a numeric comparison of problem characteristics.
\end{tcolorbox}

\begin{tcolorbox}[fonttitle=\bfseries, title=Meta-Solver Strategy, size=small]
A Meta-Solver Strategy is a combination of solvers and decisions that can be used to solve a problem instance.
Meta-Solver Strategies can use other Meta-Solver Strategies as subroutines, which allows for a hierarchical structure of Meta-Solver Strategies.
\end{tcolorbox}

Decisions are naturally expressed with if-else statements, where a condition selects which solver to run.
In our DSL, however, an else branch is mandatory.
A solver must always be executed, and we can never have a case where we do not receive a solution.
Conditions rely on problem instance properties and standard boolean expressions, enabling non-trivial Meta-Solver Strategies that select solvers based on these properties.
Because solvers always need to produce a problem solution, we can omit return statements in a function, as they would be redundant.

\begin{Verbatim}[commandchars=\\\{\}]
\textcolor{violet}{solve} \textcolor{teal}{VRP} \textcolor{darkblue}{vrp}:
  \textcolor{violet}{if} \textcolor{darkblue}{vrp}.\textcolor{darkblue}{dimension} >= 4:
    \textcolor{darkblue}{vrp}.\textcolor{brown}{LkhVrpSolver}() \textcolor{darkgreen}{// Solve large problems classically}
  \textcolor{violet}{else}:
    \textcolor{darkblue}{vrp}.\textcolor{brown}{QrispVrpSolver}() \textcolor{darkgreen}{// Otherwise, use quantum computing}
\end{Verbatim}

Meta-Solver Strategies often involve decomposition into multiple subproblems.
The DSL therefore supports nested \texttt{solve} statements to model hierarchical solving processes.
This allows arbitrary nesting depth and grants much flexibility for designing Meta-Solver Strategies.
The \texttt{foreach} loop and \texttt{TSP[]} type in Figure~\ref{fig:running-example} show that the language must support collection types and iteration structures.
Subproblems may be single instances (e.g., TSP-to-QUBO) or collections of instances (e.g., clustered TSPs).

\begin{figure}[htbp]
\begin{subfigure}{.49\textwidth}
  \centering
  \begin{Verbatim}[commandchars=\\\{\}]
  \textcolor{violet}{solve} \textcolor{teal}{VRP} \textcolor{darkblue}{vrp}:
   \textcolor{darkgreen}{// Hybrid clustering approach}
   \textcolor{darkgreen}{// for larger instances}
   \textcolor{violet}{if} \textcolor{darkblue}{vrp}.\textcolor{darkblue}{dimension} >= 4:
    \textcolor{darkblue}{vrp}.\textcolor{brown}{ClusterAndSolveVrpSolver}():
     \textcolor{violet}{solve} \textcolor{teal}{ClusterVRP} \textcolor{darkblue}{clusterVrp}:
      \textcolor{darkgreen}{// Cluster the VRP}
      \textcolor{darkblue}{clusterVrp}.\textcolor{brown}{TwoPhaseClusterer}():
       \textcolor{violet}{solve} \textcolor{teal}{TSP[]} \textcolor{darkblue}{tsps}:
        \textcolor{darkgreen}{// Now solve the TSP clusters}
        \textcolor{violet}{foreach} \textcolor{darkblue}{tsp} \textcolor{violet}{in} \textcolor{darkblue}{tsps}:
         \textcolor{violet}{if} \textcolor{darkblue}{tsp}.\textcolor{darkblue}{dimension} > 5:
          \textcolor{darkblue}{tsp}.\textcolor{brown}{LkhTspSolver}()
         \textcolor{violet}{else}:
          \textcolor{darkblue}{tsp}.\textcolor{brown}{QuboTspSolver}():
           \textcolor{violet}{solve} \textcolor{teal}{QUBO} \textcolor{darkblue}{qubo}:
            \textcolor{darkblue}{qubo}.\textcolor{brown}{QrispQuboSolver}(
             \textcolor{darkred}{"MaxNumVars" = "5"})
   \textcolor{darkgreen}{// Quantum solver for small instances}
   \textcolor{violet}{else}:
    \textcolor{darkblue}{vrp}.\textcolor{brown}{QrispVrpSolver}()
  \end{Verbatim}
  \caption{Example for the decomposition of a VRP problem into TSPs and the solving process of the TSP subproblems.}
  \label{fig:running-example}
\end{subfigure}
\hfill
\begin{subfigure}{.49\textwidth}
  \centering
  \begin{Verbatim}[commandchars=\\\{\}]
  \textcolor{darkgreen}{// Define TspViaQUBOStrategy}
  \textcolor{violet}{solve} \textcolor{teal}{TSP} \textcolor{darkblue}{tsp}:
   \textcolor{violet}{if} \textcolor{darkblue}{tsp}.\textcolor{darkblue}{dimension} > 5:
    \textcolor{darkblue}{tsp}.\textcolor{brown}{LkhTspSolver}()
   \textcolor{violet}{else}:
    \textcolor{darkblue}{tsp}.\textcolor{brown}{QuboTspSolver}():
     \textcolor{violet}{solve} \textcolor{teal}{QUBO} \textcolor{darkblue}{qubo}:
      \textcolor{darkblue}{qubo}.\textcolor{brown}{QrispQuboSolver}(
       \textcolor{darkred}{"MaxNumVars" = "5"})

  \textcolor{darkgreen}{// Use TspViaQUBOStrategy in VRP strategy}
  \textcolor{violet}{solve} \textcolor{teal}{VRP} \textcolor{darkblue}{vrp}:
   \textcolor{violet}{if} \textcolor{darkblue}{vrp}.\textcolor{darkblue}{dimension} >= 4:
    \textcolor{darkblue}{vrp}.\textcolor{brown}{ClusterAndSolveVrpSolver}():
     \textcolor{violet}{solve} \textcolor{teal}{ClusterVRP} \textcolor{darkblue}{clusterVrp}:
      \textcolor{darkblue}{clusterVrp}.\textcolor{brown}{TwoPhaseClusterer}():
       \textcolor{violet}{solve} \textcolor{teal}{TSP[]} \textcolor{darkblue}{tsps}:
        \textcolor{violet}{foreach} \textcolor{darkblue}{tsp} \textcolor{violet}{in} \textcolor{darkblue}{tsps}:
         \textcolor{darkblue}{tsp}.\textcolor{brown}{TspViaQUBOStrategy}()
   \textcolor{violet}{else}:
    \textcolor{darkblue}{vrp}.\textcolor{brown}{QrispVrpSolver}()
  \end{Verbatim}
  \caption{Example for a composition of Meta-Solver Strategies with identical logic to Figure~\ref{fig:running-example}.
  }
  \label{fig:running-example-composed}
\end{subfigure}
\end{figure}

Finally, the language needs to be able to express the composition of Meta-Solver Strategies.
Every Meta-Solver Strategy also acts as a solver for its target problem type and should therefore be callable as a solver in the language.
This way, we can improve readability of the language by reducing the amount of code to define a strategy, and promote the reuse and sharing by allowing existing strategies to serve as subroutines, as showcased in Figure~\ref{fig:running-example-composed}.

%% file: sections/04_implementation.tex
\section{Implementation}

\begin{figure}[htbp]
  \centering
  \includegraphics[width=\linewidth]{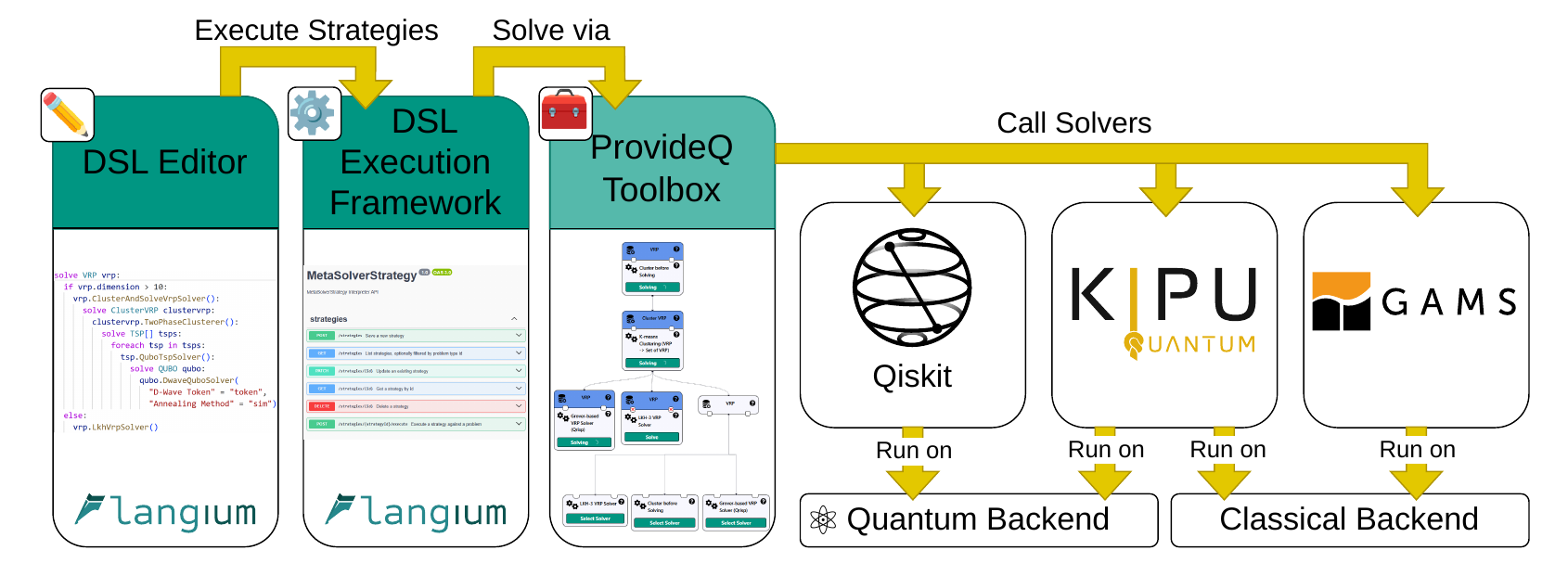}
  \caption{The technology stack of the Meta-Solver Strategy DSL.}
  \label{fig:dsl-stack}
\end{figure}

Being able to execute Meta-Solver Strategies is essential for the practical use of the DSL.
The technology stack that powers the DSL comprises multiple layers, as can be seen in Figure~\ref{fig:dsl-stack}.
First, there is the \textit{DSL Editor}, which enables the writing, editing, and saving of Meta-Solver Strategies.
Then there is the \textit{DSL Execution Framework}, which reads a Meta-Solver Strategy, and interprets it by calling the appropriate services to execute solvers.
On the third layer, there is the main service that the execution framework calls, the \textit{ProvideQ toolbox} \cite{eichhorn_provideq_2025}, which acts as the underlying execution layer for the Meta-Solver Strategies.
The ProvideQ toolbox provides many classical and quantum services, means to manage subproblem orchestration, as well as problem type definitions.
The many services that the ProvideQ toolbox uses as backends, like Qiskit~\cite{Qiskit}, the Kipu Quantum Hub~\cite{kipu_quantum_hub}, or GAMS~\cite{bussieck2004general}, make up the fourth layer of the technology stack.
Finally, the fifth layer consists of quantum and classical backends that the services call into.
The implementation can be found on GitHub: \href{https://github.com/ProvideQ/meta-solver-strategy-lang}{https://github.com/ProvideQ/meta-solver-strategy-lang}.

\subsection{DSL Editor}
The editor is implemented as a web-based application in Typescript using the Langium framework~\cite{langium}.
The DSL is defined using a context-free grammar to generate the starting point for a full-fledged backend supporting the Language Server Protocol (LSP), which only needs some customizations to support DSL specific features.
In order to do this, Langium also generates a parser and an abstract syntax tree (AST) for the DSL which can also be used to interpret a Meta-Solver Strategy in the DSL.
The editor provides a rich user experience for writing Meta-Solver Strategies, including syntax highlighting, error checking, and auto-completion.
All these smart editor features are using problem type definitions and solver definitions from the ProvideQ toolbox.
For example, it is possible to provide users with auto-completion for the solvers that are available to run on a problem instance with a given problem type.
As users write their Meta-Solver Strategy, the editor communicates with the ProvideQ toolbox to fetch the necessary information about problem types, solvers available for a problem and more, to provide a rich editing experience.
The editor also allows users to save their Meta-Solver Strategy and share them with others using a custom backend which also houses the execution framework, and can be accessed via a REST API.

\subsection{Execution Framework}
The framework is implemented in Typescript as it also uses Langium's~\cite{langium} generated AST to interpret the Meta-Solver Strategies.
When executing a Meta-Solver Strategy, the framework traverses the AST and calls the appropriate services in the ProvideQ toolbox.
A REST API request starts the execution, for which it only needs the id of a Meta-Solver Strategy saved in the framework, as well as the problem instance in the ProvideQ toolbox that the strategy should be executed for.
For this, it is first required to create a problem instance in the ProvideQ toolbox, which can be done via the REST API as well.
The problem instance has to be of the same problem type as the one that the Meta-Solver Strategy solves.

%% file: sections/05_case_study.tex
\section{Using the DSL in Practice}

In this section, we demonstrate how our DSL can be used in practice.
We show that a Meta-Solver Strategy created with our DSL can be executed via the ProvideQ Toolbox~\cite{eichhorn_provideq_2025}.
Our focus is to demonstrate the practical applicability of our DSL, rather than evaluating the performance of the Meta-Solver Strategies created with it.

\begin{figure}[htbp]
  \centering
  \includegraphics[width=0.925\linewidth]{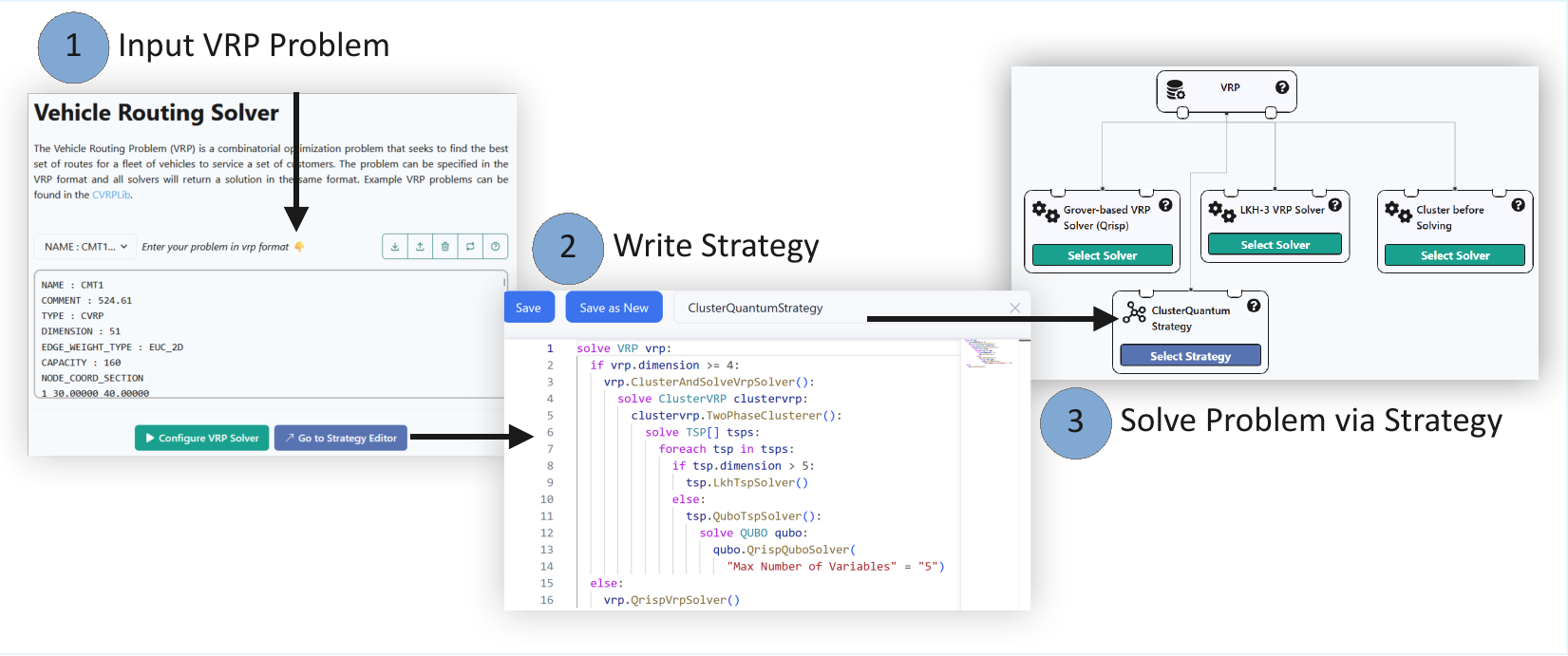}
  \caption{Workflow in the ProvideQ Toolbox to create a strategy and to use it to solve a problem instance.}
  \label{fig:workflow}
\end{figure}

In our case study, we use the Meta-Solver Strategy that we previously introduced in Figure~\ref{fig:running-example-composed}.
Although it is not meant to be the state-of-the-art for VRP solving, it serves well as a showcase.
To solve a VRP instance, the first decision checks the dimension of the problem instance, which is the number of nodes in the VRP graph.
If the dimension is smaller than 4, it uses a Qrisp quantum solver \cite{seidel_qrisp_2022} that utilizes Grover's algorithm \cite{grover_fast_1996} to solve the problem instance.
Otherwise, the problem instance is clustered into smaller TSP subproblems using a 2-Phase clustering approach \cite{feld2019hybrid} and solves each separately.
For each TSP cluster, it is checked if the dimension exceeds 5, and in this case, it uses the classical LKH-3 solver \cite{helsgaun_extension_2017}.
Smaller TSP clusters are transformed into QUBO problems that are solved with a Qrisp quantum solver using QAOA \cite{farhi_quantum_2014}.
The workflow to create and execute this Meta-Solver Strategy in the ProvideQ Toolbox is shown in Figure \ref{fig:workflow}.
We used two VRP instances with different sizes to demonstrate the execution.
P-n19-k2 by Augerat et al. \cite{augerat_computational_1995} with dimension 19 and a constructed VRP with dimension 3 that we call P-n3-k1 is very small, so it can be solved on a simulated quantum backend.

\begin{figure}[htbp]
  \centering
  \includegraphics[width=\linewidth]{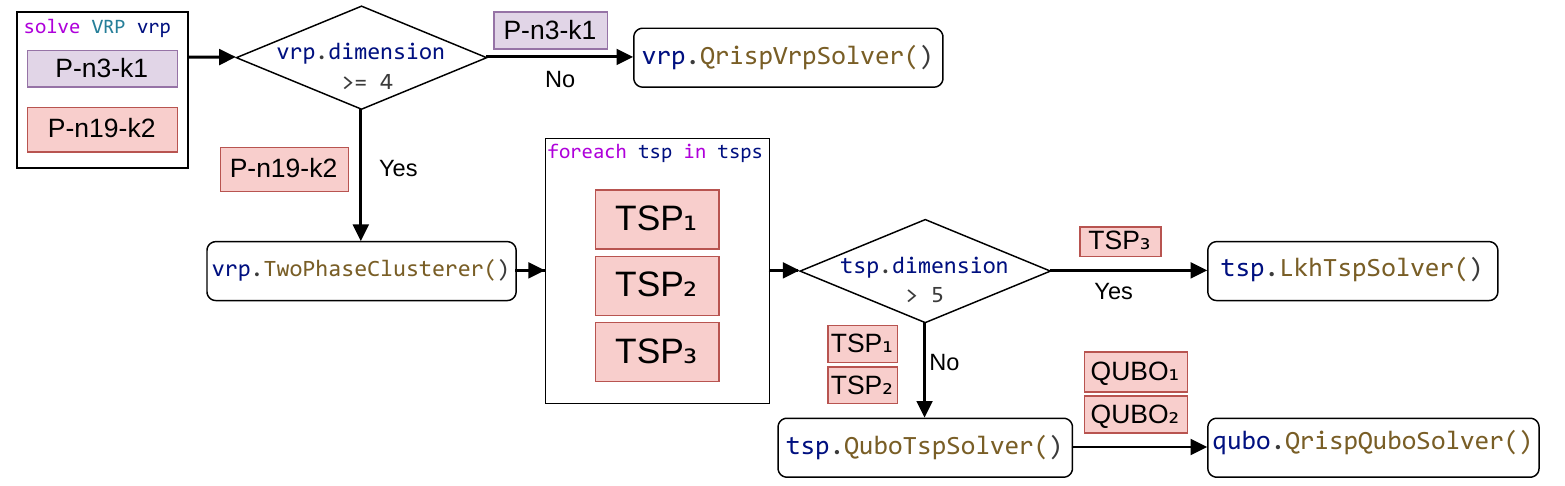}
  \caption{Solution paths for the problem instances P-n3-k1 and P-n19-k2 in the example from Figure~\ref{fig:running-example-composed}.}
  \label{fig:solution-paths}
\end{figure}

When executing the Meta-Solver Strategy for the two different problem instances, we can verify that they are executed as expected based on the different problem characteristics.
The result shows that P-n3-k1, our constructed problem with a dimension of 2, is executed directly using the Qrisp quantum solver, which is the path shown in purple in Figure \ref{fig:solution-paths}.
P-n19-k2, has a dimension of 19, and therefore exceeds the threshold of 4.
Thus, the Meta-Solver Strategy clusters the problem instance into smaller subproblems.
The resulting 3 TSP clusters with dimensions 11, 8, and 2 are then solved separately with the bigger TSPs (11 and 8) being solved using the classical LKH-3 solver, whereas the TSP of dimension 2 is transformed into a QUBO problem and solved via QAOA.

Our case study shows that the Meta-Solver Strategies created with our DSL can be executed and employed successfully to solve different problem instances based on their characteristics.
This also demonstrates the simplicity of the workflow using the ProvideQ Toolbox, which experts can use to share their Meta-Solver Strategies with others.

%% file: sections/06_conclusion.tex
\section{Conclusion and Future Work}

In this paper, we have presented a novel DSL for the definition and execution of Meta-Solver Strategies.
The DSL can express decisions to choose between quantum and classical methods to solve problems based on problem-specific characteristics.
Technical details are abstracted away, allowing users of the DSL to focus on the solving process, and decoupling solver semantics from concrete backends so Meta-Solver Strategies remain valid as toolchains evolve.
Furthermore, the DSL allows to capture expert knowledge, enabling the sharing and reuse of Meta-Solver Strategies across different problem domains.
Meta-Solver Strategies can be defined in an editor offering rich syntax support and error checking, and can be executed via an execution framework that we developed based on the ProvideQ Toolbox~\cite{eichhorn_provideq_2025}.
To demonstrate the capabilities of our DSL, we presented a case study on two VRP instances.
We found that executing the same Meta-Solver Strategies for the different VRP instances led to different solution paths, which shows that the DSL works as intended.
Future work could focus on adding support for formalized Meta-Solver Strategies, as seen in \cite{eichhorn_ensuring_2026}.
This could be implemented via automated checks in the DSL editor, enabling users to write validated strategies that are guaranteed to yield correct results.
Another avenue for future work could be the implementation of additional execution environments for the DSL aside from the ProvideQ Toolbox.
Next steps could also include the implementation of benchmarking support to compare different strategies across a range of problem instances, and to analyze the performance of strategies in terms of solution quality, runtime, and cost.